\documentstyle[prc,aps,preprint]{revtex}
\begin{document}
\draft
\title{Energy loss, electron screening and the astrophysical
$^3$He(d,p)$^4$He cross section }
\author{K. Langanke$^1$, T.~D.~Shoppa$^1$, C.~A. Barnes$^1$ and C. Rolfs$^2$}
\address{$^1$ W.~K.~Kellogg Radiation Laboratory, 106-38\\
California Institute of Technology\\
Pasadena, CA 91125 USA\\
$^2$ Institut f\"ur Physik mit Ionenstrahlen\\
Ruhr-Universit\"at Bochum\\
D-44780 Bochum, Germany}
\date{\today}
\maketitle

\begin{abstract}
We reanalyze the low-energy $^3$He(d,p)$^4$He cross section measurements
of Engstler {\it et al.} using recently measured energy loss data
for proton and deuteron beams in a helium gas. 
Although the new $^3$He(d,p)$^4$He
astrophysical S-factors are significantly 
lower than those reported by Engstler {\it et al.},
they clearly show the presence of electron screening effects.
From the new astrophysical S-factors we find an electron screening energy
in agreement with the adiabatic limit.
\end{abstract}


The penetration through the Coulomb barrier forces the (non-resonant)
cross section $\sigma(E)$ between charged particles 
to drop exponentially with decreasing energy $E$.
(Energies are in the center-of-mass system throughout this paper.)
As a consequence, the cross section at the very low energies at which stellar
hydrostatic burning takes place is in most cases too small to be measured
directly in the laboratory. It is therefore customary in nuclear
astrophysics to measure cross sections to energies
as low as possible and then to extrapolate the data to the energy
appropriate for the astrophysical application. Conventionally this
extrapolation is performed in terms of the astrophysical S-factor defined 
(in a model-independent way) by
\begin{equation}
\sigma(E) = {S(E) \over E} \exp (-2\pi\eta(E))\;,
\end{equation}
where
$\eta(E) = Z_1 Z_2 e^2 \sqrt{ \mu/E }$
is the Sommerfeld parameter for
initial nuclei of charges $Z_1, Z_2$ and reduced
mass $\mu$. The exponential Gamow-factor in Eq.~(1) describes the 
s-wave penetration
through the Coulomb barrier of point-like charges and thus accounts
for the dominant energy dependence of the cross section at energies far below
the Coulomb barrier. Additional energy dependences due to nuclear structure,
strong interaction, phase space, finite nuclear size, etc. 
are expected to leave the S(E)-factor
a slowly varying function of energy
for non-resonant reactions. 

It is common strategy in nuclear astrophysics to reduce the uncertainties
related to the extrapolation of the S(E)-factor by pushing laboratory
measurements to even lower energies. However, as has been 
pointed out by Assenbaum
{\it et al.} \cite {Assenbaum}, there is a 
potential problem with this strategy as, at
the lowest energies now accessible in laboratory experiments, the electrons
present in the target (and possibly also in the projectile) may lead
to an enhancement of the measured cross section over the desired cross
section for bare nuclei by partially screening 
the Coulomb barrier between projectile
and target. As discussed in \cite{Assenbaum}, the screening effect is equivalent
to giving the colliding nuclei an extra attraction (described by an energy
increment $U_e$). Thus, the nuclei may be considered as tunneling
 the Coulomb barrier at an 
effective incident energy $E_{\rm eff}=E+U_e$. 
The resulting enhancement of the measured
cross section $\sigma_{\rm exp} (E)$
over the cross section for bare nuclei $\sigma_{\rm bn} (E)$ can 
then be defined as
\begin{equation}
f(E) = \sigma_{\rm exp} (E) / \sigma_{\rm bn} (E) =
\sigma (E + U_e) / \sigma (E).
\end{equation}
Considering that $U_e << E$ at those
energies currently accessible in experiments and approximating
$S(E)/E = S(E_{\rm eff})/E_{\rm eff}$
one finds 
\cite{Assenbaum}
\begin{equation}
f(E) 
\approx \exp \left\{ \pi \eta(E) \frac{U_e}{E} \right\}.
\end{equation}
In general, the screening energy $U_e$ is a function of energy.
However, it has become  customary to express the enhancement of measured cross
sections due to electron screening in terms of a {\it constant}
screening energy \cite{NMR}. For atomic (deuteron and helium) targets
this assumption has been justified in \cite{Shoppa} for the energies
at which screening effects are relevant.

Experimentally, electron screening effects have been established and
studied intensively by the M\"unster/Bochum group 
\cite{NMR,Engstler,Angulo,Greife,Prati}. 
By fitting the expression 
for the enhancement factor $f(E)$, as given in Eq.~(3), to the ratio
of measured cross section over the expected bare-nuclear cross section
($\sigma_{\rm bn}$ is usually derived by extrapolating cross sections from
higher energies where screening effects are negligible), electron
screening energies $U_e$ have been derived for several nuclear reactions 
\cite{Engstler,Angulo,Greife,Prati}. 
Surprisingly, these screening energies 
have been reported to be larger than the
adiabatic limit in which the electrons adjust instantaneously to the
change in nuclear configuration, and in which it is assumed that the
associated gain in electron binding energy is entirely transferred to
the relative motion of the colliding nuclei. As long as it is justified
to treat the nuclei as infinitely heavy, which appears to be a valid
approximation at the energies involved, the adiabatic limit should constitute
the maximum screening energy possible.

The most pronounced excess of the experimentally derived screening energy
over the adiabatic limit has been reported for the 
$^3$He(d,p)$^4$He
reaction \cite{Prati}. With an atomic $^3$He gas target, cross sections
were measured down to $E=5.88$ keV. At this energy,
the observed cross section exceeds the 
extrapolated bare nuclear cross section by about 50$\%$.
Furthermore, the enhancement of the data over the bare nuclear cross section
fits the expected exponential energy dependence with a screening energy
of $U_e=186\pm9$ eV \cite{Prati}. This value is significantly larger than
the adiabatic limit of $U_e=120$ eV \cite{Shoppa}. Note that one possible
source of uncertainty is the extrapolation of the bare-nuclei
cross section. For
$^3$He(d,p)$^4$He,
the extrapolation appears to be sufficiently well under control. For example,
the parametrization of the available data for energies $E=40$ keV to 10 MeV
predicts an astrophysical S-factor in the relevant energy regime $E=5-40$ keV which agrees
very well with the one calculated in a microscopic cluster model
\cite{Bluge}.

Obviously the determination of electron screening effects require high-precision
measurements. In particular, the effective energy in the target or,
equivalently,
the energy loss in any matter upstream of the 
target has to be known very precisely. In
\cite{Engstler,Prati}, the authors used the stopping power data 
for deuterons in helium as
tabulated in \cite{Ziegler}. These tables were derived by extrapolation
of the stopping power for deuterons above 100 keV to lower energies,
assuming a linear dependence on the projectile velocity
\cite{Lindhard,Firsov}. As noted by Lindhard \cite{Lindhard1}
and by Bang \cite{Bang}, this extrapolation can contain substantial errors. In fact,
recent measurements of the stopping power of low-energy 
protons and deuterons
($\leq 25$ keV) in a helium gas \cite{stopping} found significantly lower values than
tabulated in \cite{Ziegler}. These smaller stopping powers are in good
agreement with a more recent calculation, based on a coupled-channel solution
for the time-dependent Schr\"odinger equation for a hydrogen beam traversing
a helium gas \cite{Grande}. In this calculation, Grande and Schiwietz
show that the stopping power at low energies is dominated by electron capture
by the projectile.
This process, however, requires a 
substantial minimum energy transfer which results
in a 
considerably reduced stopping power at lower energies than would be
expected from a
velocity-proportional extrapolation of data from higher energies.

We now re-derive the low-energy
$^3$He(d,p)$^4$He
astrophysical 
S-factors for the Engstler {\it et al.} measurements \cite{Engstler}
using the stopping power data of \cite{stopping} rather than the
tabulated values of \cite{Ziegler}; the latter values were adopted
in \cite{Engstler} and in the recent reanalysis of the data by Prati 
{\it et al.} \cite{Prati}. We translated the stopping power data
of \cite{stopping} into energy losses as a function of deuteron+$^3$He
(c.m.) energies $E$ and then fitted these data by a smooth curve. The resulting
energy loss functions are shown in Fig. 1 for the two different
pressures (0.1 Torr and 0.2 Torr) at which the experiment \cite{Engstler}
has been performed. For comparison, the energy loss function as derived from
the Ziegler-Andersen table \cite{Ziegler}
is also shown. From Fig. 1 we observe that at
the lowest energy ($E=5.88$ keV), at which Engstler {\it et al.} report
$^3$He(d,p)$^4$He astrophysical
S-factors (0.2 Torr), the measured energy loss \cite{stopping}
is about 80 eV less than the tabulated value. At $E=10$ keV, the difference
is still 48 eV. Note that these differences are significant compared with
of the energy previously attributed 
to electron screening ($U_e=186$ eV). In fact,
using the reduced energy losses will result in reduced astrophysical
S-factors.
Correspondingly we expect the electron
screening energy deduced from the data to decrease.

To derive
$^3$He(d,p)$^4$He astrophysical
S-factors for the new energy loss data, we first transformed
the S(E)-factors into cross sections, using the S(E) data and energies $E$
as given in Table 1 of Ref. \cite{Engstler}; a $3.8\%$ intrinsic error
has been added in quadrature to the data, in accordance with Ref. \cite{Prati}.
Then we derived new
effective energies $E'=E+\Delta E_{\rm loss}$, where $\Delta E_{\rm loss}$
is the excess of the tabulated energy \cite{Ziegler}
losses over the recently measured values \cite{stopping}.
The cross section data, now attributed to the effective energy $E'$,
are then transformed into astrophysical S-factors S$(E')$. For the
exponent in the Gamow factor we used $2\pi\eta(E)=68.75/\sqrt(E)$ 
(with $E$ in keV),
in accordance with \cite{Engstler,Chulick}. As expected, the new 
astrophysical S-factors
are significantly smaller than those reported in \cite{Engstler,Prati}
(Fig. 2). 

Following Refs. \cite{Engstler,Prati}
we then determined the electron screening energy $U_e$ by
fitting expression (3) to the ratio of the new S(E) data to the bare-nuclei
astrophysical S-factors. As in \cite{Prati}
we used the 3-rd order polynomial
($n=3$) parametrization given in \cite{Chulick} for the bare-nuclei
cross section. 
We find $U_e=117\pm7$ eV 
with a $\chi^2$-value of 0.5 per degree of freedom. To roughly estimate
the uncertainty of the extrapolated bare-nuclei cross section on $U_e$
we have repeated the fit for the 
4-th order polynomial ($n=4$) parametrization of \cite{Chulick}.
We then find $U_e=134\pm8$ eV ($\chi^2=0.4$). Both values are in agreement
within uncertainties
with the adiabatic limit, which has been shown to apply at the low
collision energies studied here \cite{Shoppa}. Thus, the 
$^3$He(d,p)$^4$He astrophysical
S-factors derived with the recently measured energy loss data do not show
the excess of screening energy reported in \cite{Prati}. We 
stress that 
the uncertainties related to the extrapolation of the bare-nuclei
cross sections are significantly larger than the statistical errors
of the experimental data,
even in a case where the extrapolation appears to be reasonably well under
control.

Note that the approximation $S(E_{\rm eff})/ E_{\rm eff}= S(E)/ E$, 
used to derive Eq. (3) from
the definition of the enhancement factor (2), is incorrect by about
$3\%$ at the lowest energies studied here, leading to an approximately
$10\%$ underestimation of the screening energy. We have therefore repeated
the determination of the screening energy, now using Eq. (2). We then find
$U_e=130\pm8$ eV and 149$\pm$9 eV for the $n=3$ and $n=4$ parametrization
of the bare-nuclei cross sections, respectively. Although these
values are slightly larger than the adiabatic screening limit,
they do not provide evidence for an excess of screening
beyond the uncertainties in the experiments involved,
as deduced
in \cite{Prati}. 

In summary, we have shown that conclusions drawn 
previously about electron screening
effects on low-energy fusion data depend very sensitively  on the assumed
energy loss in the target and in matter up-stream of the target, 
for which only rather scarce data
exist at such low collision energies. Recent measurements \cite{stopping} and
theoretical work \cite{Grande} indicate that the energy loss of a hydrogen
beam transversing a helium gas is significantly less than given by the
standard tables. If these reduced energy losses are applied to the
$^3$He(d,p)$^4$He astrophysical
S-factors, we have found an electron screening energy 
in agreement with the theoretically expected adiabatic limit ($U_e=120$ eV),
within uncertainties, that no
longer requires an
unexplained screening excess. Our work clearly
stresses the need for improved low-energy stopping power data for this and other
reactions in which an excess of the screening energy over the adiabatic
limit has been reported \cite{NMR}. 
Further work on low-energy stopping powers in gas and
solid targets has already been initiated
\cite{prop}.
If
this work confirms the reduced stopping powers at very low energy, 
it will
also validate the general strategy in nuclear astrophysics to achieve
more reliable astrophysical
nuclear cross sections by steadily lowering
the energies at which the cross sections are measured in the
laboratory, as the electron screening effects, at least for atomic targets,
can then be considered to be understood.

The work was supported in part by the National Science Foundation,
Grant Nos. PHY94-12818 and PHY94-20470, and by the Deutsche 
Forschungsgemeinschaft.
One of the authors (KL) would like to thank Professor Jens Lindhard
for a valuable discussion.

\begin{figure}
\caption{Comparison of the measured
energy losses for a hydrogen beam traversing a helium gas target
\protect\cite{stopping} with tabulated values \protect\cite{Ziegler}.
All energies are measured in the center-of-mass system of the colliding nuclei.
The energy losses have been calculated for the two pressures
($\rho$=0.2 Torr and 0.1 Torr) at which the experiments of Ref.
\protect\cite{Engstler} have been performed.
}
\end{figure}

\begin{figure}
\caption{Comparison of the astrophysical
S-factors for the $^3$He(d,p)$^4$He reaction as reported in
\protect\cite{Engstler} (open symbols) and as extracted using the
revised energy losses (filled symbols). The squares and triangles refer
to the data measured at 0.2 Torr and 0.1 Torr, respectively.
The solid curve shows the 3rd order polynomial ($n=3$)
parametrization of the bare-nuclei cross section
taken from \protect\cite{Chulick}. The dashed curve represents the best fit
to the new astrophysical
S-factor data using Eq. (2) and a constant screening energy
of $U_e=130$ eV. 
}
\end{figure}

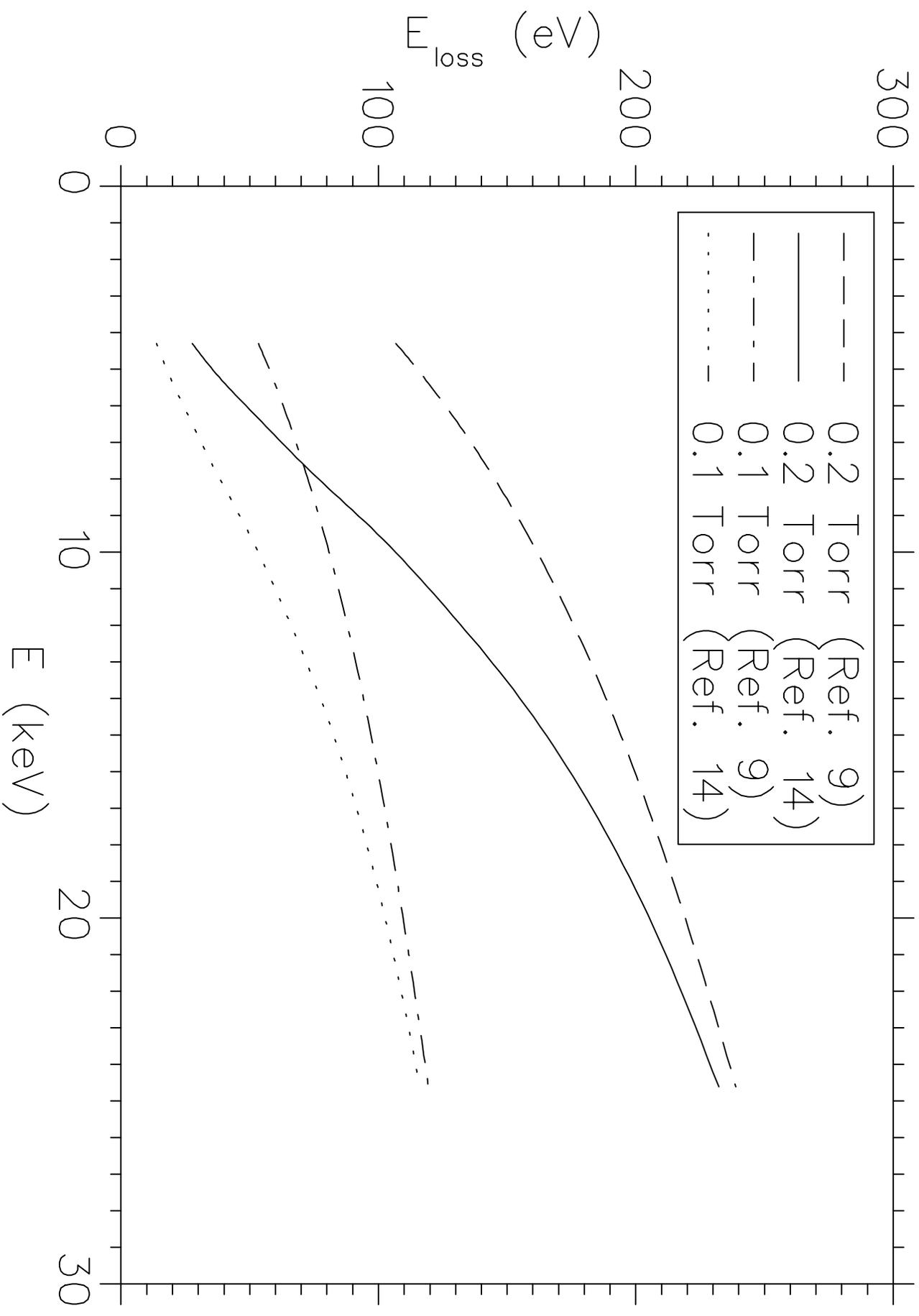
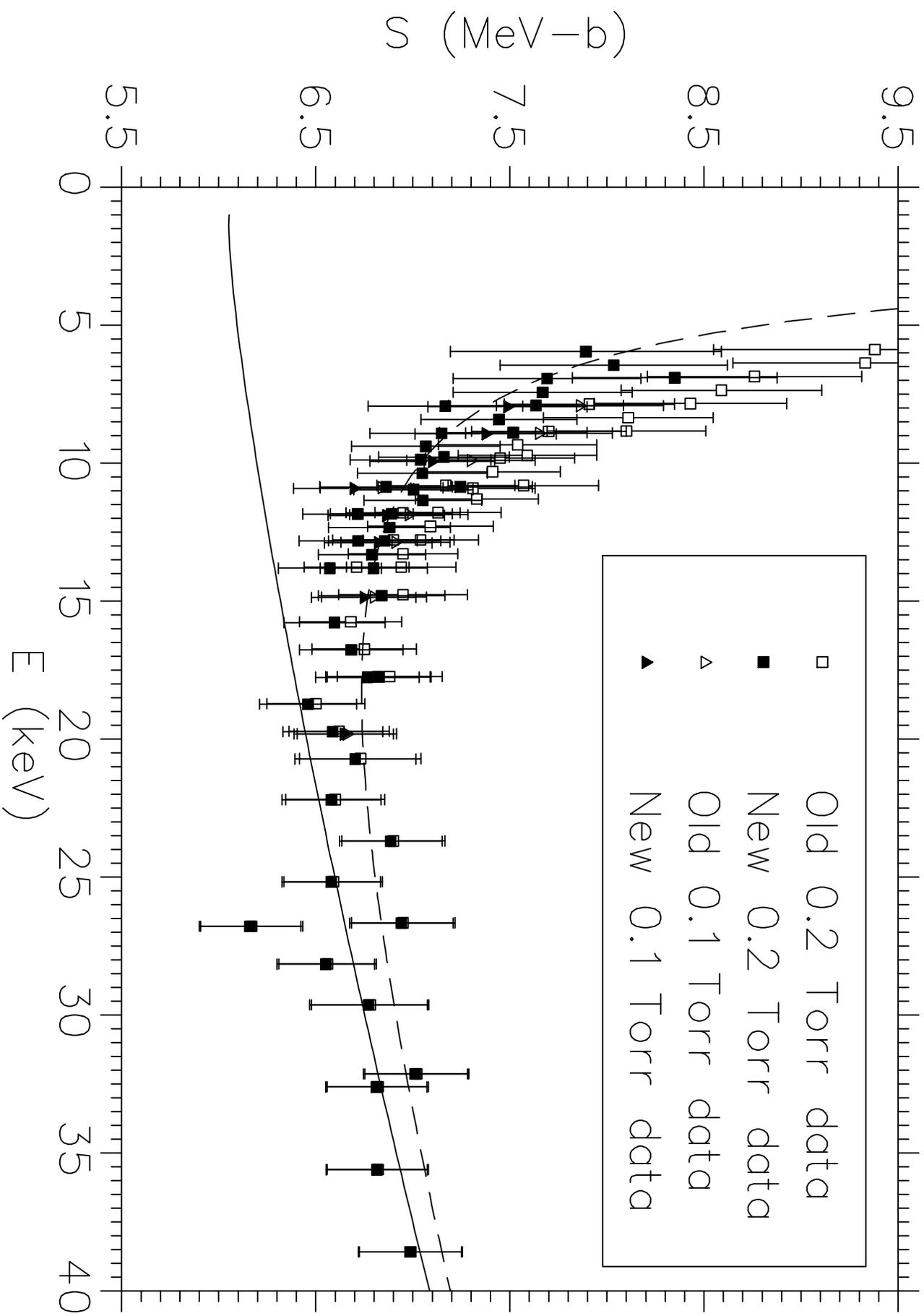
\end{document}